\documentclass[a4paper,twocolumn,10pt,noarxiv,accepted=2023-02-02]{quantumarticle}
\pdfoutput=1
\usepackage{amsmath}
\usepackage{amssymb}
\usepackage{hyperref}
\usepackage[numbers,sort&compress]{natbib}
\usepackage[utf8]{inputenc}
\usepackage[english]{babel}
\usepackage[T1]{fontenc}
\usepackage{amsmath}
\usepackage{amssymb}
\usepackage{hyperref}
\usepackage{graphicx}
\usepackage{dcolumn}
\usepackage{bm}
\usepackage{braket}
\usepackage{tabularx}
\usepackage{threeparttable}

\hypersetup{colorlinks=true,citecolor=magenta,urlcolor=magenta,linkcolor=magenta}

\begin{document}

\title{Performance analysis of quantum repeaters enabled by deterministically generated photonic graph states}

\author{Yuan Zhan\textsuperscript{1}}
\orcid{0000-0001-5450-8879}
\author{Paul Hilaire\textsuperscript{2}}
\orcid{0000-0002-7144-6953}
\author{Edwin Barnes\textsuperscript{2}}
\orcid{0000-0003-1666-9385}
\author{Sophia E. Economou\textsuperscript{2}}
\orcid{0000-0002-1939-5589}
\author{Shuo Sun\textsuperscript{1,*}}
\orcid{0000-0003-4171-0466}

\affiliation{\textsuperscript{1} JILA and Department of Physics, University of Colorado, Boulder, Colorado 80309, USA
\\
\textsuperscript{2}\, Department of Physics, Virginia Tech, Blacksburg, Virginia 24061, USA
\\
\textsuperscript{*}\, Correspondence should be addressed to: shuosun@colorado.edu
}

% \author{Yuan Zhan}
% \orcid{0000-0001-5450-8879}
% \affiliation{JILA and Department of Physics, University of Colorado, Boulder, Colorado 80309, USA}
% \author{Paul Hilaire}
% \orcid{0000-0002-7144-6953}
% \affiliation{Department of Physics, Virginia Tech, Blacksburg, Virginia 24061, USA}
% \author{Edwin Barnes}
% \orcid{0000-0003-1666-9385}
% \affiliation{Department of Physics, Virginia Tech, Blacksburg, Virginia 24061, USA}
% \author{Sophia E. Economou}
% \orcid{0000-0002-1939-5589}
% \affiliation{Department of Physics, Virginia Tech, Blacksburg, Virginia 24061, USA}
% \author{Shuo Sun}
% \orcid{0000-0003-4171-0466}
% \affiliation{JILA and Department of Physics, University of Colorado, Boulder, Colorado 80309, USA}
% \email{shuosun@colorado.edu}

% \date{\today}

\begin{abstract}
By encoding logical qubits into specific types of photonic graph states, one can realize quantum repeaters that enable fast entanglement distribution rates approaching classical communication. However, the generation of these photonic graph states requires a formidable resource overhead using traditional approaches based on linear optics. Overcoming this challenge, a number of new schemes have been proposed that employ quantum emitters to deterministically generate photonic graph states. Although these schemes have the potential to significantly reduce the resource cost, a systematic comparison of the repeater performance among different encodings and different generation schemes is lacking. Here, we quantitatively analyze the performance of quantum repeaters based on two different graph states, i.e. the tree graph states and the repeater graph states. For both states, we compare the performance between two generation schemes, one based on a single quantum emitter coupled to ancillary matter qubits, and one based on a single quantum emitter coupled to a delayed feedback. We identify the numerically optimal scheme at different system parameters. Our analysis provides a clear guideline on the selection of the generation scheme for graph-state-based quantum repeaters, and lays out the parameter requirements for future experimental realizations of different schemes.
\end{abstract}

\maketitle

\section{Introduction}
\label{intro}

The central question in quantum networking is how to faithfully transmit a quantum signal over a noisy and lossy channel~\cite{Kimble:2008uv,Northup:2014tt,Stephanie:2018uf}. Overcoming this challenge requires intermediate quantum nodes, referred to as quantum repeaters, to refresh the entanglement that is degraded due to channel loss and decoherence~\cite{PhysRevLett.81.5932}. Standard proposals for quantum repeaters rely on heralded entanglement generation for remote entanglement~\cite{Duan:2001tt,RevModPhys.83.33}; these protocols require two-way classical signaling over the entire chain of quantum repeaters to transmit one qubit. Such signaling implies that the requisite quantum memory coherence time must be substantially longer than the end-to-end communication times, a demanding requirement in experimental realizations. In addition, the two-way signaling substantially limits the achievable communication rate.

More recently, quantum repeaters based on photonic graph states were proposed to overcome these challenges~\cite{Azuma:2015vq,PhysRevX.10.021071,PhysRevA.104.052623,Zhang:22}. This approach employs quantum encoding to correct photon losses and mitigate operation errors, eliminating the need for long-lived quantum memories. In addition, graph-state-based quantum repeaters permit high entanglement rates that are close to classical communication~\cite{Hilaire2021resource,PhysRevX.10.021071}. Despite their great promise, the generation of such photonic graph states requires a formidable resource overhead using conventional approaches based on linear optics and measurement-based feedforward~\cite{PhysRevX.5.041007,PhysRevA.95.012304}.

To overcome this challenge, several groups have proposed schemes for deterministic generation of different types of photonic graph states~\cite{PhysRevLett.103.113602,PhysRevLett.105.093601,Pichler:2017wl,PhysRevX.7.041023,PhysRevLett.125.223601,PhysRevA.104.013703,Michaels2021multidimensional,PRXQuantum.2.040345,Li:2022tw,shapourian2022modular}, including the tree graph and repeater graph states that are considered as resource states for quantum repeaters~\cite{PhysRevX.7.041023,PhysRevLett.125.223601,Li:2022tw,shapourian2022modular}. These schemes employ a single spin-tagged quantum emitter to sequentially emit entangled photons, inspired by the original proposal from Lindner and Rudolph~\cite{PhysRevLett.103.113602}. While single quantum emitters can only generate photons entangled in one dimension, which can be efficiently represented by matrix product states~\cite{PhysRevLett.95.110503,Schwartz:2016tn,Thomas:2022tr}, it is possible to generate entangled photons in higher dimensions such as tree graph and repeater graph states by coupling the quantum emitter to either ancillary matter qubits~\cite{PhysRevX.7.041023,Li:2022tw} or a delayed feedback~\cite{PhysRevLett.125.223601}. Quantitative performance analyses have also been performed for the ancillary-qubit-assisted generation scheme in the realization of graph-state-based quantum repeaters~\cite{PhysRevX.10.021071,Hilaire2021resource}. However, the performance of the scheme based on delayed feedback has not been analyzed yet. More importantly, it remains unclear which scheme offers better performance in the realization of a graph-state-based quantum repeater.

\begin{figure*}[tbh]
\centering
\includegraphics[width=1.75\columnwidth]{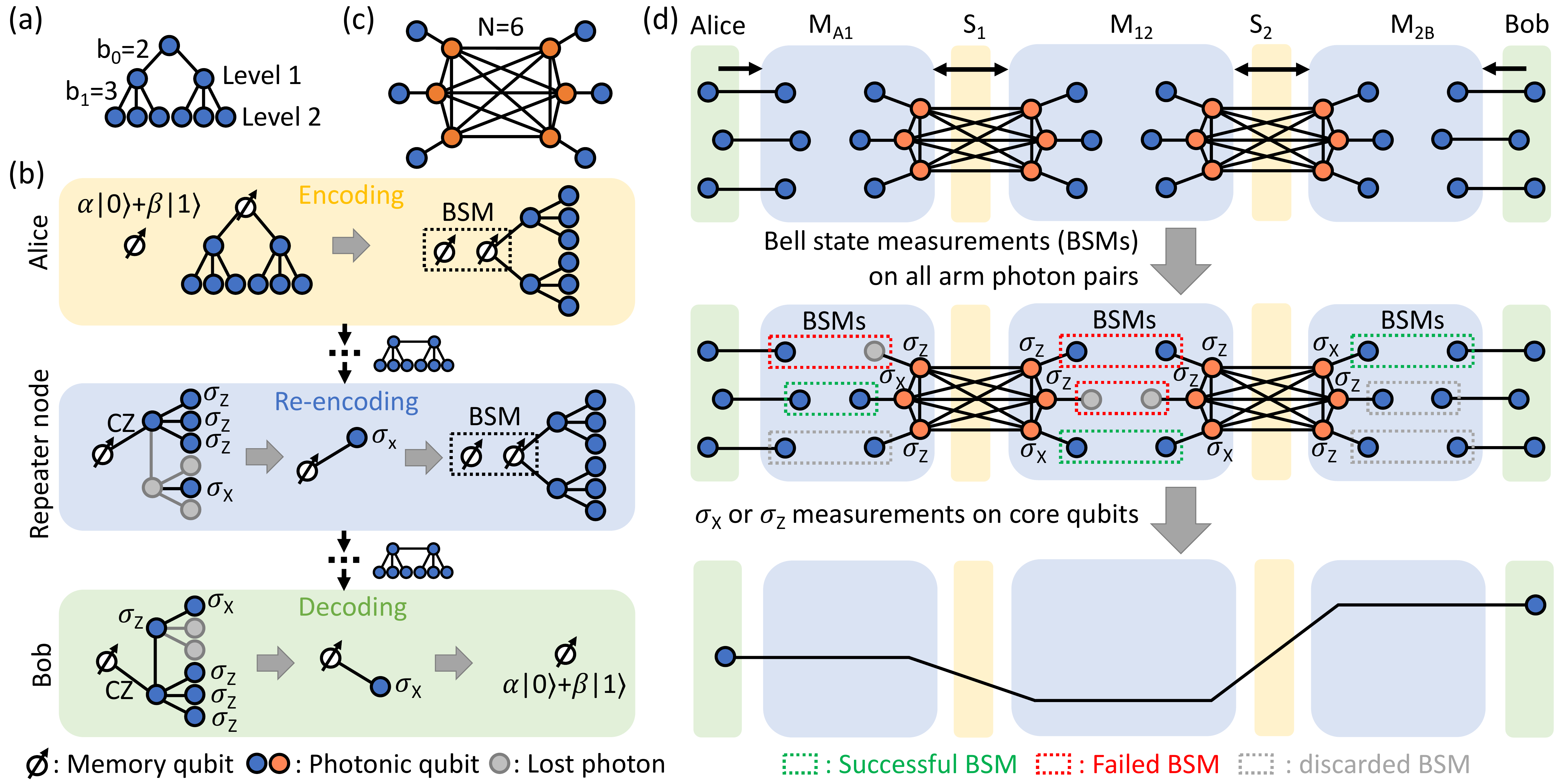}
\caption{(a) An example graph representing a tree graph state with branching parameters $\{b_0=2,b_1=3\}$. (b) Schematic of the quantum repeater protocol based on tree graph states. Alice (yellow block) encodes a logical qubit into a tree graph state via a Bell state measurement on the quantum memory and the root qubit of the tree, and sends the photons to the first repeater node. Each repeater node (blue block) decodes the logical qubit from the arriving photons and re-encodes it into a complete tree graph state. Bob (green block) retrieves the logical qubit into his local quantum memory. (c) An example of the repeater graph with $N=6$ core nodes (orange circles) and $N=6$ arm nodes (blue circles). (d) Schematic of the quantum repeater protocol based on repeater graph states. To establish an entanglement link between Alice and Bob (green blocks), each source node $S_i$ (yellow blocks) generates a repeater graph state and sends half of the photons to the left measurement node $M_{i-1,i}$ (blue blocks), and the other half to the right one $M_{i,i+1}$. At each measurement node, a Bell state measurement is performed on each pair of arm photons, and a single-qubit measurement in either the $\sigma_X$ or $\sigma_Z$ basis is performed on each core qubit for either entanglement swapping or graph detachment, depending on the success or failure of the corresponding Bell state measurement. An entangled Bell pair between Alice and Bob is then established. BSM, Bell state measurement; CZ, controlled-$Z$ gate; $\sigma_{X,Y,Z}$, Pauli $X$, $Y$, or $Z$ matrices.}
\label{fig1}
\end{figure*}

In this paper, we quantitatively compare the performance of the two generation schemes in the realization of two different graph-state-based quantum repeater protocols, and identify the numerically optimal scheme for different system parameters. Our analysis provides a clear guideline on the selection of the generation scheme for graph-state-based quantum repeaters, and lays out the parameter requirements for future experimental realizations of different schemes.

This paper is organized as follows. Section~\hyperref[review]{II} reviews the quantum repeater protocols based on photonic graph states and the schemes for the generation of the required photonic graph states. Section~\hyperref[reff]{III} defines the figure of merit of the repeater performance by considering a specific application of quantum key distribution. Section~\hyperref[comparison]{IV} quantitatively evaluates and compares the performance and resource costs of both generation schemes in the realization of different graph-state-based quantum repeater protocols. Section~\hyperref[conclusion]{V} discusses the implications of our results in experimental realizations of graph-state-based quantum repeaters.

\section{Review of existing graph-state-based quantum repeater protocols and their implementation schemes}
\label{review}

\begin{figure*}[tbh]
\centering
\includegraphics[width=2.07\columnwidth]{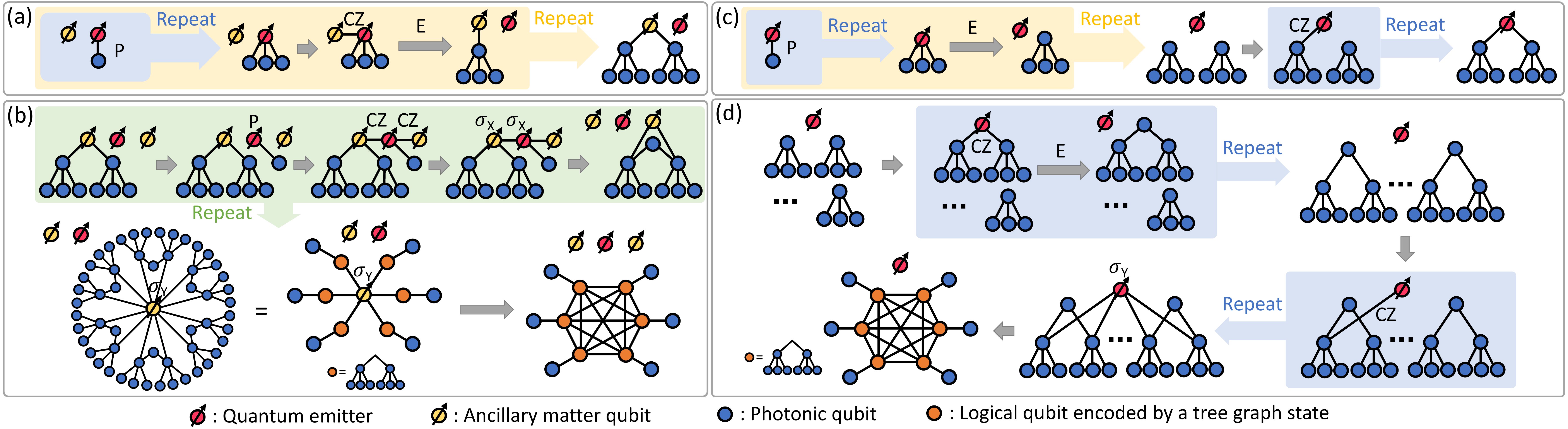}
\caption{(a)(b) Generation protocols for the tree graph state (a) and the repeater graph state (b) using a single quantum emitter with the assistance of a few ancillary matter qubits~\cite{PhysRevX.7.041023,Li:2022tw}. (c)(d) Generation flows of the tree graph state (c) and the repeater graph state (d) using a single quantum emitter with the assistance of a delayed feedback. This scheme is slightly modified from the original proposal~\cite{PhysRevLett.125.223601} to improve the generation speed and the error robustness (see Appendix~\ref{appendixA}). P, the $P$ gate which emits a photon entangled with the emitter spin; E, the $E$ gate which emits a photon that carries the state of the emitter spin and detaches the emitter spin from the graph; CZ, the controlled-$Z$ gate either between two matter qubits or between a matter qubit and a photon.}
\label{fig2}
\end{figure*}{}

\subsection{Graph-state-based quantum repeater protocols}
\label{protocol}

Two pioneering protocols have been proposed for quantum repeaters based on photonic graph states. We first review the protocol based on tree graph states. Figure~\ref{fig1}(a) shows a graph used to represent a specific tree graph state. The tree graph can be described by a set of branching parameters $\{b_i\}$, where $i\in \{0,1,\dots,d-1\}$ is the level index in the tree graph (the root node has an index of $i=0$), $d$ is the total depth of the tree graph, and $b_i$ is the number of leaf nodes connected with each level-$i$ node. Figure~\ref{fig1}(b) illustrates the quantum repeater protocol based on tree graph states, as proposed in Ref.~\cite{PhysRevX.10.021071}. The sender, Alice, encodes a logical qubit into a tree graph state consisting of many physical photons. The repeater node may receive only a fraction of the physical photons due to photon losses. However, as long as a complete subset of photons required for loss correction is present~\cite{PhysRevLett.97.120501}, the repeater node can retrieve the logical qubit and re-encode it into another complete tree. This is based on the loss tolerance property of the tree graph state~\cite{PhysRevLett.97.120501}. The logical qubit is eventually transmitted to Bob through all the repeater nodes, and Bob simply needs to retrieve this logical qubit into his local quantum memory.

We now review the second protocol based on repeater graph states. Figure~\ref{fig1}(c) shows an example graph representing a repeater graph state. A repeater graph state consists of $N$ core nodes [orange circles in Fig.~\ref{fig1}(c)] and $N$ arm nodes [blue circles in Fig.~\ref{fig1}(c)]. The $N$ core nodes comprise a fully connected graph, and each arm node is connected to a core node. Figure~\ref{fig1}(d) shows the schematic of the quantum repeater protocol based on repeater graph states, as proposed in Ref.~\cite{Azuma:2015vq}. The two users of the quantum network, Alice and Bob, are separated by an array of source and measurement nodes. Each source node $S_i$ generates a repeater graph state and sends half of the photons to the left measurement node $M_{i-1,i}$, while the other half go to the right one $M_{i,i+1}$. Bell state measurements are performed at each measurement node $M_{i, i+1}$ between each pair of arm photons from the source nodes $S_i$ and $S_{i+1}$. A successful Bell state measurement on one of these pairs creates an entanglement link between two corresponding core photons, one from the source node $S_i$ and the other from $S_{i+1}$. An entangled Bell pair between Alice and Bob can then be established by repeating this procedure in each measurement node, followed by certain single-qubit measurements on each core photon. To boost the overall success probability and entanglement fidelity, one can replace each core photon with a tree-encoded logical qubit to make the protocol more robust to photon losses and logical errors~\cite{Azuma:2015vq,Hilaire2021resource}.

We note that both quantum repeater protocols discussed above can enable fast entanglement distribution rates. The protocol repetition rate is solely determined by the graph state generation time and measurement time at each local node~\cite{Hilaire2021resource}. Therefore, the characteristic time of these two quantum repeater protocols accords with the ``third-generation quantum repeaters'' that require only one-way signaling~\cite{Muralidharan:2016tq}, unlike the conventional two-way repeaters where the entanglement rate is limited by the round trip time of the two-way classical signaling.

For convenience, in the rest of this article, we use ``tree repeater'' and ``RGS repeater'' to refer to the two different graph-state-based quantum repeater protocols discussed above.

\subsection{Schemes for deterministic generation of tree graph and repeater graph states}
\label{scheme}

Standard approaches for generating tree graph and repeater graph states rely on probabilistic fusion gates realized by linear optics, which requires a formidable resource overhead~\cite{PhysRevX.5.041007,PhysRevA.95.012304}. To overcome this challenge, several schemes have been proposed to deterministically generate the required photonic graph states~\cite{PhysRevX.7.041023,PhysRevLett.125.223601,Li:2022tw}. These schemes utilize a single quantum emitter to sequentially emit photons that are entangled into a one-dimensional chain, which can be entangled into the desired tree graph or repeater graph state with the assistance of either additional ancillary matter qubits~\cite{PhysRevX.7.041023,Li:2022tw} or a delayed feedback~\cite{PhysRevLett.125.223601}. Figures~\ref{fig2}(a) and~\ref{fig2}(b) show the detailed gate sequences to generate tree graph and repeater graph states, respectively, using a single quantum emitter coupled with a few ancillary matter qubits~\cite{PhysRevX.7.041023}. Three elementary operational gates are required for this scheme: the $P$ gate, corresponding to emission of a photon that is entangled with the internal spin qubit of the emitter; the $E$ gate, corresponding to emission of a photon that carries the state of the internal spin qubit of the emitter, by which the emitter spin is detached from the graph; and the $CZ$ gate, which applies a controlled-$Z$ operation between the internal spin qubit of the emitter and an ancillary matter qubit. Figures~\ref{fig2}(c) and~\ref{fig2}(d) show the detailed gate sequences to generate tree graph states and repeater graph states, respectively, using a single quantum emitter coupled with a delayed feedback~\cite{PhysRevLett.125.223601}. The elementary gates in this scheme are similar to the ancillary-qubit-assisted scheme, except that the $CZ$ gate here is on the internal spin qubit of the emitter and a photon, which can be realized by a scattering process of a single photon from the emitter~\cite{PhysRevLett.96.153601,Fan:2010ti}. We refer the readers to Refs.~\cite{PhysRevX.7.041023} and~\cite{PhysRevLett.125.223601} for detailed physical processes that realize the aforementioned elementary gates in the two schemes, respectively.

We note that neither aforementioned generation scheme generates photons in an ideal order for measurements in either repeater protocols. It has been shown in Ref.~\cite{Li:2022tw} that the ancillary-qubit-assisted scheme can be slightly modified to generate photons in the graph state in a desired order at the expense of using more matter qubits. However, without the help of additional resources, a delay line is necessary to correct the photon arrival order for measurements. This delay line will result in additional photon losses, which has not been taken into account in previous performance analyses. Our analysis, as presented below, will be the first to take into account the effect of such additional delay lines.

\section{Figure of merit for repeater performance evaluation}
\label{reff}

To compare the performance of the two generation schemes in the two different repeater protocols, we consider the specific application of quantum key distribution. The results obtained here will also serve as a qualitative comparison of the performance of the two generation schemes in other applications of graph-state-based quantum repeaters. Similar to the analysis performed in Ref.~\cite{PhysRevX.10.021071}, we use the effective secret key rate ($R_{\text{eff}}$) as the measure of the repeater performance in quantum key distribution, defined as the maximally achievable rate of secret keys that can be shared between Alice and Bob per unit resource overhead and attenuation length, given by
\begin{equation}
\begin{aligned}
R_{\text{eff}}=rP_{\text{succ}}\frac{1}{T_{\text{graph}}}\frac{1}{mn}\frac{L}{L_{\text{att}}},
\end{aligned}
\label{eq1}
\end{equation}
where $r$ is the secret fraction, defined as the ratio between the number of post-extracted secure keys and the number of initially shared raw keys in the asymptotic limit of infinitely long keys~\cite{RevModPhys.81.1301}, $P_{\text{succ}}$ is the probability of Bob successfully receiving a logical qubit sent from Alice in the case of the tree repeater, or Alice and Bob successfully establishing an entangled Bell state in the case of the RGS repeater, $T_{\text{graph}}$ is the time it takes for each repeater node to generate the required photonic graph state, $m$ is the number of evenly-separated repeater nodes between Alice and Bob, $n$ is the number of matter qubits needed at each repeater node, $L$ is the distance between Alice and Bob, and $L_{\text{att}}$ is the attenuation length of an optical photon in the communication channel, defined as the length at which the transmission probability of a photon decays to $1/e$.

In Eq.~\ref{eq1}, the secret fraction $r$ depends on the specific quantum key distribution protocol. Here, we focus on a specific protocol, known as the six-state protocol~\cite{PhysRevLett.81.3018,10.5555/2011333.2011337}. For this protocol, the secret fraction $r$ can be calculated analytically under the assumption of one-way classical post-processing and perfect classical error correction~\cite{RevModPhys.81.1301}, given by
\begin{equation}
\begin{aligned}
r=F-h\left(1-F\right)-Fh\left(\frac{3F-1}{2F}\right),
\end{aligned}
\label{eq2}
\end{equation}
where $h(x)=1-x\log_2(x)-(1-x)\log_2(x)$ is the binary entropy function, and $F$ is the quantum state fidelity of the transmitted qubit in the case of the tree repeater, or the quantum state fidelity of the shared entangled Bell state between the sender and the receiver in the case of the RGS repeater. We note that $r$ goes to zero when $F\lesssim 87.4\%$, in which case no unconditionally secure keys can be extracted. In Appendix~\ref{appendixB}, we derive the fidelity $F$ under realistic errors including the spin coherence time $t_{\text{coh}}$ and the depolarization probability of each photon in the communication channel $\varepsilon_\text{depol}$. Other operations such as the single-qubit gates and measurements on the emitter spin qubit, the $CZ$ gate between two matter qubits, and the Bell state measurements on two matter or photonic qubits are assumed to be error-free.

In Eq.~\ref{eq1}, the success probability $P_{\text{succ}}$ depends on the repeater protocol. For the tree repeater based on a tree graph state with branching parameters $\{b_i\}$ (defined in Sec.~\ref{review}), the success probability $P_{\text{succ}}^{\text{tree}}$ is given by~\cite{PhysRevLett.97.120501,PhysRevX.10.021071}
\begin{widetext}
%\begin{equation}
\begin{align}
P_{\text{succ}}^{\text{tree}}=\left\{\left[(1-\mu+\mu R_1)^{b_0}-(\mu R_1)^{b_0}\right]\left(1-\mu+\mu R_2\right)^{b_1}\right\}^{m+1},
\end{align}
\label{eq3}
%\end{equation}
\end{widetext}
where $\mu$ is the single-photon loss probability between neighboring repeater nodes, which can be expressed as
\begin{equation}
\begin{aligned}
\mu=1-(1-\mu_{\text{ext}})(1-\mu_{\text{coup}})(1-\mu_{\text{int}})(1-\mu_{\text{del}}),
\end{aligned}
\label{eq4}
\end{equation}
where $\mu_{\text{ext}}$, $\mu_{\text{coup}}$, $\mu_{\text{int}}$, and $\mu_{\text{del}}$ are the probabilities that a single photon is lost in the communication channel, at the coupling between the source and the communication channel, inside the source during the generation process, and in the delay line for photon arrival order correction, respectively. In Eq.~\ref{eq3}, $R_i$ is the probability of obtaining an outcome from an indirect $\sigma_Z$ measurement on any given photon found in the $i$-th level of the tree graph state, given by~\cite{PhysRevLett.97.120501}
\begin{equation}
\begin{aligned}
R_i=1-\left[1-(1-\mu)\left(1-\mu+\mu R_{i+2}\right)^{b_{i+1}}\right]^{b_i}.
\end{aligned}
\label{eq5}
\end{equation}

For the RGS repeater, the success probability $P_{\text{succ}}^{\text{RGS}}$ is given by~\cite{Azuma:2015vq,Hilaire2021resource}
\begin{equation}
\begin{aligned}
P_{\text{succ}}^{\text{RGS}}=\left[\left(1-\left(1-P_{\text{BSM}}\right)^{N/2}\right)P_{\sigma_X}^2P_{\sigma_Z}^{N-2}\right]^{m+1},
\end{aligned}
\label{eq6}
\end{equation}
where $N$ is the number of branches in the repeater graph state (defined in Sec.~\ref{review}), $P_{\text{BSM}}$ is the probability of a successful photonic Bell state measurement, and $P_{\sigma_X}$ and $P_{\sigma_Z}$ are the probabilities of obtaining a result from the single-qubit $\sigma_X$ and $\sigma_Z$ measurements on a tree-encoded core qubit, respectively. $P_{\text{BSM}}$, $P_{\sigma_X}$ and $P_{\sigma_Z}$ are given by~\cite{Azuma:2015vq,Hilaire2021resource}
\begin{equation}
\begin{aligned}
&P_{\text{BSM}}=\frac{(1-\mu)^2}{2}, \\
&P_{\sigma_X}=R_0, \\
&P_{\sigma_Z}=(1-\mu+\mu R_1)^{b_0},
\end{aligned}
\label{eq7}
\end{equation}
where $\mu$ is the single-photon loss probability between a source node and its nearest measurement node, which can be given by the same form as Eq.~\ref{eq4}. In Eq.~\ref{eq7}, $b_i$ is the branching parameter of the encoding tree graph state of each core logical qubit in the repeater graph state, and $R_i$ is the probability of obtaining an outcome from an indirect $\sigma_Z$ measurement on any photon in the $i$-th level of the encoding tree graph state of a core logical qubit, which can be calculated by Eq.~\ref{eq5}.

In Eq.~\ref{eq1}, the graph state generation time $T_{\text{graph}}$ is given by~\cite{PhysRevX.10.021071,Hilaire2021resource}
\begin{widetext}
%\\begin{equation}
\begin{align}
&T_{\text{tree}}^{\text{ancilla}}\approx \prod_{i=0}^{d-1}b_it_P^a+\left(\beta b_0+\sum_{l=1}^{d-2}\prod_{i=0}^{l}b_i\right)t_E^a+\sum_{l=0}^{d-2}\prod_{i=0}^{l}b_it_{\text{CZ}}^a, \notag \\
&T_{\text{tree}}^{\text{feedback}}\approx \prod_{i=0}^{d-1}b_it_P^f+\sum_{l=0}^{d-2}\prod_{i=0}^{l}b_it_E^f+\sum_{l=0}^{d-2}\prod_{i=0}^{l}b_it_{\text{CZ}}^f, \notag \\
&T_{\text{RGS}}^{\text{ancilla}}\approx N\left[\left(1+\prod_{i=0}^{d-1}b_i\right)t_P^a+\sum_{l=0}^{d-2}\prod_{i=0}^{l}b_it_E^a+\left(2+\sum_{l=0}^{d-2}\prod_{i=0}^{l}b_i\right)t_{\text{CZ}}^a+2t_M\right]+t_M, \notag \\
&T_{\text{RGS}}^{\text{feedback}}\approx N\left[\prod_{i=0}^{d-1}b_it_P^f+\left(\frac{1}{\beta}+\sum_{l=0}^{d-2}\prod_{i=0}^{l}b_i\right)t_E^f+\left(b_0+\sum_{l=0}^{d-2}\prod_{i=0}^{l}b_i\right)t_{\text{CZ}}^f\right]+t_M,
\label{eq8}
\end{align}
%\end{equation}
\end{widetext}
where in the expressions for $T_{\text{tree}}^{\text{ancilla}}$ and $T_{\text{tree}}^{\text{feedback}}$, $b_i$ and $d$ are the branching parameters and the depth of the tree graph state, while in the expressions for $T_{\text{RGS}}^{\text{ancilla}}$ and $T_{\text{RGS}}^{\text{feedback}}$, $b_i$ and $d$ are the branching parameters and the depth of the encoding tree in the repeater graph state. The parameters $t_P$, $t_E$, $t_{\text{CZ}}$ with superscripts $a$ or $f$ are the operation times of a $P$ gate, an $E$ gate, and a $CZ$ gate in the ancillary-qubit- or feedback-assisted generation scheme, respectively. $t_M$ is the time for a single-qubit measurement on a spin. The application of ancillary-qubit-assisted scheme in the tree repeater protocol, along with the application of the feedback-assisted scheme in both repeater protocols, requires part of the photons in the graph state to have longer wavepackets than other photons in order to boost the fidelity of the spin-photon $CZ$ gate. Therefore, we define $\beta$ as the ratio of the wavepacket length between these longer photons and the rest of the photons. This parameter $\beta$ naturally appears in the expression of $T_{\text{tree}}^{\text{ancilla}}$, because in the tree repeater protocol, a spin-photon $CZ$ gate is required for photons situated in the first level of the tree graph state in the re-encoding step at each repeater node~\cite{PhysRevX.10.021071}. In the calculations of $T_{\text{tree}}^{\text{feedback}}$ and $T_{\text{RGS}}^{\text{feedback}}$, the parameter $\beta$ implicitly appears in the ratio of $t_E^f$ and $t_P^f$, since each photon, except for the arm photons, generated by an $E$ gate needs to scatter with the emitter for the spin-photon $CZ$ gate in the feedback-assisted generation scheme.

\section{Repeater performance analysis and comparison}
\label{comparison}

\begin{figure}[t]
\centering
\includegraphics[width=0.97\columnwidth]{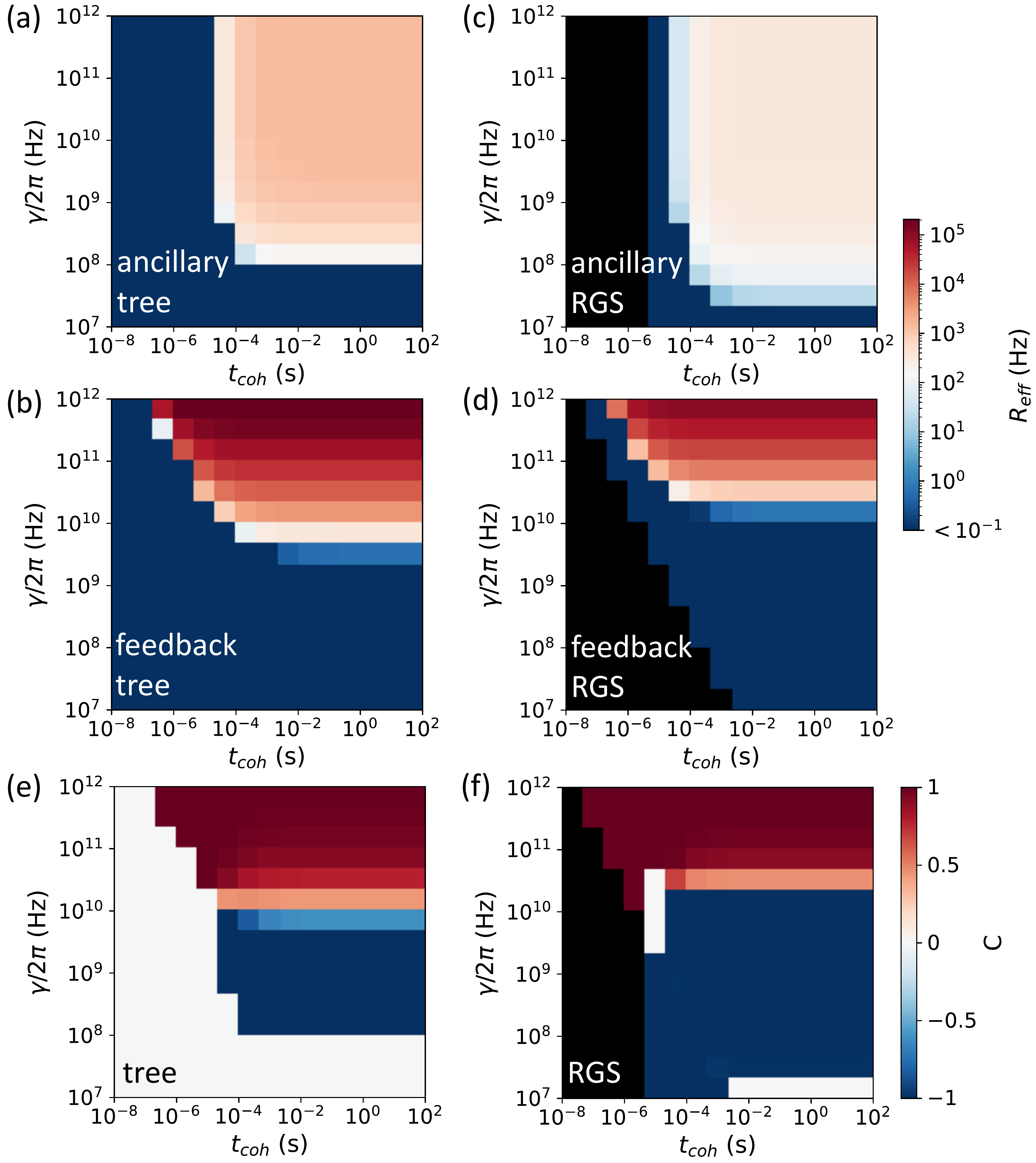}
\caption{(a)-(d) The maximally achievable effective secret key rate $R_{\text{eff}}$ as a function of the optical linewidth $\gamma$ and the spin coherence time $t_\text{coh}$ of the quantum emitter for a (a) tree repeater network realized by the ancillary-qubit-assisted generation scheme, (b) tree repeater network realized by the feedback-assisted-generation scheme, (c) RGS repeater network realized by the ancillary-qubit-assisted generation scheme, and (d) RGS repeater network realized by the feedback-assisted-generation scheme. The black blocks indicate that it is impossible to extract any unconditionally secure keys using any photonic graph states for the specific values of $\gamma$ and $t_\text{coh}$. (e)(f) Normalized difference of $R_{\text{eff}}$ between the two generation schemes for tree repeater (e) and RGS repeater (f) protocols. The white blocks indicate a very close performance between the two schemes, which typically happens when both schemes have a very poor performance (i.e. $R_{\text{eff}}$ is close to zero), thus giving a close-to-zero difference of $R_{\text{eff}}$.}
\label{fig3}
\end{figure}

We now quantitatively analyze and compare the performance of the two generation schemes in the realization of both graph-state-based quantum repeater protocols. Two important parameters in the generation of the tree graph and repeater graph states are the optical linewidth, $\gamma$, and the spin coherence time, $t_\text{coh}$, of the quantum emitter. The optical linewidth determines the minimum time it takes to implement the $P$ and $E$ gates in both generation schemes, and the spin-photon $CZ$ gate in the feedback-assisted scheme. The spin coherence time, roughly speaking, limits the maximum time one can take to generate the required graph state. Ideally, we would like to implement both generation schemes with a quantum emitter that possesses a large optical linewidth (which can be Purcell enhanced) and a long spin coherence time, but these two properties may not be achieved simultaneously by the same atomic system. Figure~\ref{fig3} compares the maximally achievable effective secret key rate $R_\text{eff}$ as a function of $\gamma$ and $t_\text{coh}$ for both the ancillary-qubit- and feedback-assisted generation schemes in the realization of both tree and RGS repeater protocols. In these calculations, we fix the rest of the parameters using realistically achievable or reasonably approximated values listed in Table~\ref{tab1}. The maximally achievable effective secret key rate at each set of $\gamma$ and $t_\text{coh}$ is calculated by optimizing over all possible photonic graph states and the number of repeater nodes. If it is impossible to extract any unconditionally secure keys using any photonic graph states, the effective secret key rate for the specific parameters $\gamma$ and $t_\text{coh}$ is depicted as a black block in the figure. This corresponds to the regime where the emitter spin coherence time is short, resulting in a low fidelity $F$ and hence a negative secret fraction $r$. In Appendix~\ref{appendixF}, we list the numerically optimized graph state shape, the number of repeater nodes, and the length of the feedback or delay line for several parameters $\gamma$ and $t_\text{coh}$ of interest.

\begin{table*}[p]
\caption{Values of fixed parameters in the calculation of the effective secret key rate $R_{\text{eff}}$ in Fig.~\ref{fig3}.}
\begin{threeparttable}
%\begin{ruledtabular}
\begin{tabularx}{2.0\columnwidth}{p{0.35\columnwidth}p{0.65\columnwidth}p{0.35\columnwidth}p{0.65\columnwidth}}
\hline\hline
Parameter &Value &Parameter &Value \\
\hline
$\mu_{\text{ext,tree}}$\tnote{ a} &$1-\exp{\left[-\frac{L}{(m+1)L_{\text{att}}}\right]}$ &$t_E^a$ &$t_P^a+t_H+t_M$\tnote{ k} \\
$\mu_{\text{ext,RGS}}$\tnote{ b} &$1-\exp{\left[-\frac{L}{2(m+1)L_{\text{att}}}\right]}$ &$t_H$ &100 ps\tnote{ l} \\
$\mu_{\text{coup}}$ &0.05\tnote{ c} &$t_M$ &10$t_P^{a/f}$\tnote{ m} \\
$\mu_{\text{int,feedback}}$\tnote{ d} &$1-\exp{\left(-\frac{n_\text{feedback}L_{\text{feedback}}}{L_{\text{att}}}\right)}$\tnote{ e} &$t_E^f$ &$\beta t_P^f+t_H+t_M$\tnote{ n} \\
$\mu_{\text{int,ancilla}}$\tnote{ f} &0\tnote{ g} &$t_\text{CZ}^a$ &100 ns\tnote{ o} \\
$\mu_{\text{del}}$ &$1-\exp{\left(-\frac{L_{\text{delay}}}{L_{\text{att}}}\right)}$\tnote{ h} &$t_\text{CZ}^f$ &$t_E^f$\tnote{ p} \\
$\varepsilon_\text{depol}$ &$5\times 10^{-5}$ &$t_\text{coh,ancilla}$\tnote{ q} &$\infty$\tnote{ r} \\
$\beta$ &500\tnote{ i} &$L$ &1000 km \\
$t_P^a$, $t_P^f$ &$1/\gamma$\tnote{ j} &$L_\text{att}$ &20 km \\
\hline\hline
\end{tabularx}
%\end{ruledtabular}
\begin{tablenotes}
\footnotesize
\item[a] $\mu_{\text{ext}}$ for the tree repeater protocol.
\item[b] $\mu_{\text{ext}}$ for the RGS repeater protocol, where a measurement node is situated in between two source nodes.
\item[c] Both cavity and waveguide quantum electrodynamics devices have shown $>99$\% efficiency from a single quantum emitter to a single photonic mode~\cite{Najer:2019to,Bhaskar:2020uj,Kindem:2020vz,Raha:2020te}. Coupling from a waveguide or a cavity mode to a single-mode fiber can achieve an efficiency $>97$\% as well~\cite{Tiecke:15}. We therefore estimate a total internal coupling loss of $5$\%.
\item[d] $\mu_{\text{int}}$ for the feedback-assisted scheme.
\item[e] In the feedback-assisted scheme, some photons need to travel in the feedback line before they can be sent to the communication channel. Therefore there is a loss due to the absorption in the feedback line depending on how many round trips the photon needs to travel in the entire feedback line, $n_\text{feedback}$, which can be 0, 1, or 2. Here, $L_{\text{feedback}}$ is the length of the feedback line for generating a specific graph state (see Appendix~\ref{appendixC} for its explicit expression). We assume the quantum efficiency of the emitter to be 1.
\item[f] $\mu_{\text{int}}$ for the ancillary-qubit-assisted scheme.
\item[g] In the ancillary-qubit-assisted scheme, photons emitted by the emitter are directly sent to the communication channel. Assuming the quantum efficiency of the emitter to be 1, the internal loss in the generation process would be zero.
\item[h] $L_{\text{delay}}$ is the length of the delay line for reordering generated photons in the graph state for measurements. It depends on both the generation scheme and the repeater protocol (see Appendix~\ref{appendixD} for its explicit expression).
\item[i] See Appendix~\ref{appendixE} for justification of this value.
\item[j] $t_P^a$ and $t_P^f$ can be taken as the pulse width of a single-photon wavepacket generated by a $P$ gate in both generation schemes, which is approximately the inverse of the optical linewidth of the emitter.
\item[k] In the ancillary-qubit-assisted scheme, an $E$ gate can be realized by a $P$ gate followed by a Hadamard gate applied on the emitter spin qubit, a Hadamard gate on the newly emitted photon, plus a single-qubit $\sigma_Z$ measurement of the spin qubit of the quantum emitter~\cite{PhysRevX.7.041023}. Since the time of a photonic Hadamard gate is generally much shorter than a spin Hadamard gate, we can express $t_E^a$ as $t_E^a=t_P^a+t_H+t_M$, where $t_H$ is the operation time of a spin Hadamard gate.
\item[l] A spin Hadamard gate can be implemented via a two-photon Raman process within $\sim 100$ ps, as has been demonstrated experimentally in self-assembled quantum dots~\cite{Press:2008tz,De-Greve:2011wq}, nitrogen-vacancy (NV) centers~\cite{Bassett:2014uq} and silicon-vacancy (SiV) centers~\cite{Becker:2016vk} in diamond.
\item[m] A projective spin measurement is typically achieved via resonance fluorescence through a spin-dependent cycling transition. Assuming the spontaneous emission time of the cycling transition is identical to the $P$ gate time (since both involve the time it takes to emit a photon), and that it takes 10 cycles to obtain a high-fidelity spin measurement result, we can estimate that $t_M=10t_P^{a/f}$, where $t_P^{a/f}$ is the $P$ gate time in either generation scheme.
\item[n] In the feedback-assisted scheme, a fraction of the photons emitted by the emitter needs to have a longer wavepacket in order to boost the fidelity of the spin-photon $CZ$ gate. These photons are generated via $E$ gates instead of $P$ gates. Therefore, the photon emission process in the $E$ gate takes a much longer time than $t_P^f$. Similar to $t_E^a$, a Hadamard gate and a spin measurement following the photon emission are typically required to implement the $E$ gate. We thus define $t_E^f=\beta t_P^f+t_H+t_M$, and we will take the value of $\beta$ from the left column of this table.
\item[o] Entanglement of two spin qubits can be realized through their local dipolar interactions, which are typically weak, ranging from hundreds of kHz to a few hundreds MHz. For example, the hyperfine interaction strength between an electron spin of a nitrogen-vacancy (NV) center in diamond and a nearby $^{13}$C or $^{14}$N nuclear spin is around 10 MHz~\cite{doi:10.1126/science.1139831,Pfaff:2013ue}, and the hyperfine interaction strength in a silicon-vacancy (SiV) center in diamond is a few MHz between an electron spin and a carbon-13 nuclear spin~\cite{PhysRevLett.123.183602} and a few hundreds MHz between an electron spin and a $^{29}$Si nuclear spin~\cite{doi:10.1126/science.add9771}. Composite microwave pulses can be employed to realize high-fidelity entangling gates, resulting in a gate time of at least hundreds of nanoseconds~\cite{Rong:2015tx}. In neutral atom systems, such local interactions can be realized by Rydberg blockade, whose strength can be tens of MHz with two atoms a few micrometers away from each other~\cite{Gaetan:2009vv}. Two-photon processes can be used to generate entanglement in hundreds of nanoseconds~\cite{PhysRevLett.104.010502}. Therefore, we assume the $CZ$ gate time to be 100 ns.
\item[p] In the feedback-assisted scheme, the $CZ$ gate time is determined by the wavepacket length of a physical photon emitted by an $E$ gate, which is approximately the operation time of an $E$ gate.
\item[q] The coherence time of the ancillary qubits in the ancillary-qubit-assisted scheme.
\item[r] Since our main focus is on the properties of the quantum emitter itself, we set the coherence time of the ancillary matter qubit to be infinity so that the ancillary matter qubit does not limit the quantum repeater performance in any way. This assumption may lead to an overestimate of the performance of the quantum repeaters realized by the ancillary-qubit-assisted generation scheme.
\end{tablenotes}
\end{threeparttable}
\label{tab1}
\end{table*}

We first compare the two generation schemes in the application of the tree repeater protocol [Figs.~\ref{fig3}(a) and~\ref{fig3}(b)]. In both figures, we observe that the maximally achievable effective key rate $R_\text{eff}$ increases as we increase the optical linewidth $\gamma$ and the spin coherence time $t_\text{coh}$. However, for the ancillary-qubit-assisted scheme, $R_\text{eff}$ saturates when $\gamma/2\pi>1$ GHz. This is because at a large enough optical linewidth, the generation time of the tree graph state will be limited by the $CZ$ gate time between the spin of the quantum emitter and the ancillary matter qubit, and thus further increasing $\gamma$ does not help to achieve a higher $R_\text{eff}$. In contrast, for the feedback-assisted scheme, $R_\text{eff}$ shows no saturation as we increase $\gamma$. This is because both the single-photon emission time and the spin-photon $CZ$ gate time are reduced when we have a larger operation bandwidth $\gamma$. For the same reason, $R_\text{eff}$ shows extremely poor values for the feedback-assisted scheme when $\gamma/2\pi<2$ GHz, regardless of the spin coherence time. For both schemes, $R_\text{eff}$ shows a saturation when we increase the spin coherence time $t_\text{coh}$. This is because when the spin coherence time $t_\text{coh}$ is long enough, the single-photon logical error caused by the spin decoherence during the photonic graph state generation will become negligible compared with the photon depolarization error in the communication channel.

To highlight the difference between the two generation schemes in the implementation of the tree repeater protocol, we calculate the normalized difference of the maximally achievable effective key rate, defined as $C=\frac{R_\text{eff,f}-R_\text{eff,a}}{R_\text{eff,f}+R_\text{eff,a}}$ where $R_\text{eff,f}$ and $R_\text{eff,a}$ are the maximally achievable effective key rates for the feedback- and ancillary-qubit-assisted schemes, respectively. A positive value in the normalized difference means that the feedback-assisted scheme performs better, whereas a negative value indicates that the ancillary-qubit-assisted scheme wins out. Figure~\ref{fig3}(e) shows the normalized difference as a function of $\gamma$ and $t_\text{coh}$. The feedback-assisted scheme generally wins out in the large optical linewidth regime, regardless of the spin coherence time, whereas the ancillary-qubit-assisted scheme generally performs better when the emitter has a small optical linewidth but a long spin coherence time. The performance of the two generation schemes in the RGS repeater protocol [Figs.~\ref{fig3}(c)(d)(f)] shows a very similar behavior as well.

\begin{figure}[t]
\centering
\includegraphics[width=1.0\columnwidth]{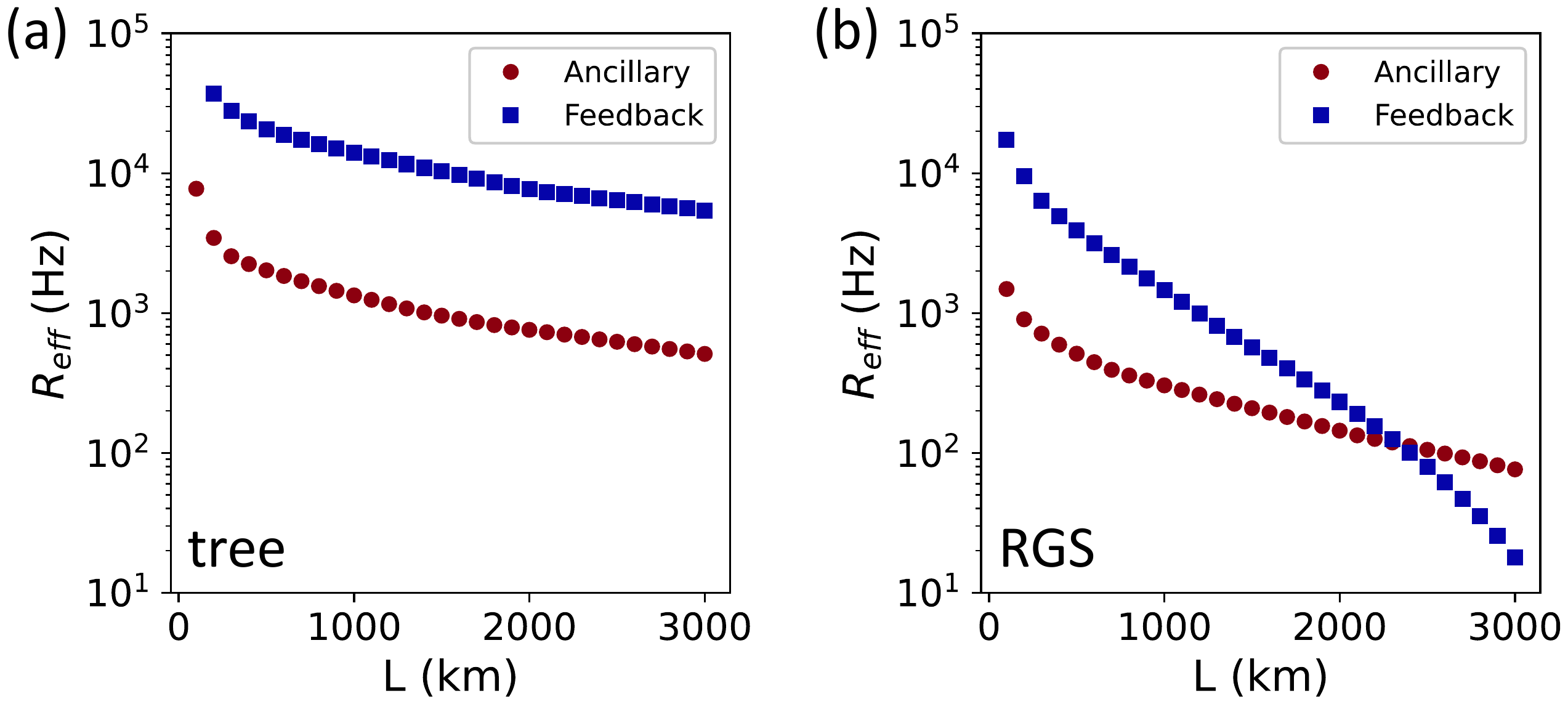}
\caption{The maximally achievable effective secret key rate $R_{\text{eff}}$ as a function of total distance between Alice and Bob, $L$, using the ancillary-qubit-assisted scheme (red dots) and feedback-assisted scheme (blue squares) for the tree repeater (a) and the RGS repeater (b). Here, we assume $\gamma/2\pi=10$ GHz and $t_{\text{coh}}=1$ ms.}
\label{fig4}
\end{figure}

So far, our analysis is limited to a fixed distance between Alice and Bob ($L=1000$ km). To understand how the difference of the two generation schemes extends into other distance regimes, we calculate the maximally achievable effective key rate $R_\text{eff}$ as a function of the total distance $L$. Figures~\ref{fig4}(a) and~\ref{fig4}(b) show the relation between $R_\text{eff}$ and $L$ for the tree repeater and RGS repeater protocols, respectively. In this calculation, we fix $\gamma/2\pi=10$ GHz and $t_{\text{coh}}=1$ ms, where the feedback-assisted scheme outperforms the ancillary-qubit-assisted scheme by about 10 times for the tree repeater and 4 times for the RGS repeater at $L=1000$ km. As shown in Fig.~\ref{fig4}, the performance of the two generation schemes shows similar distance dependence when used to implement the tree repeater protocol, but quite different when used to implement the RGS repeater protocol. Specifically, when used to implement the RGS repeater protocol, the performance of the ancillary-qubit-assisted generation scheme decays much slower than the feedback-assisted one, and starts to outperform the feedback-assisted scheme when $L>2400$ km. This result can be understood from the following two observations. First, the ancillary-qubit-assisted scheme tends to outperform the feedback-assisted one when used to generate a large-size graph state. A large-size graph state is generally required for a longer distance between neighboring repeater nodes, which is preferred for a longer end-to-end distance to achieve a better scaling with the end-to-end distance. This performance difference in generating a large graph state is due to the additional internal photon loss arising in the feedback-assisted scheme when we extend the length of the feedback line in order to generate a large-size photonic graph state. Second, the tree repeater protocol generally requires a graph state of a much smaller size than the RGS repeater protocol, which explains why the relative performance of the two generation schemes does not show much difference in the distance dependence for the tree repeater protocol.

\begin{figure}[t]
\centering
\includegraphics[width=1.0\columnwidth]{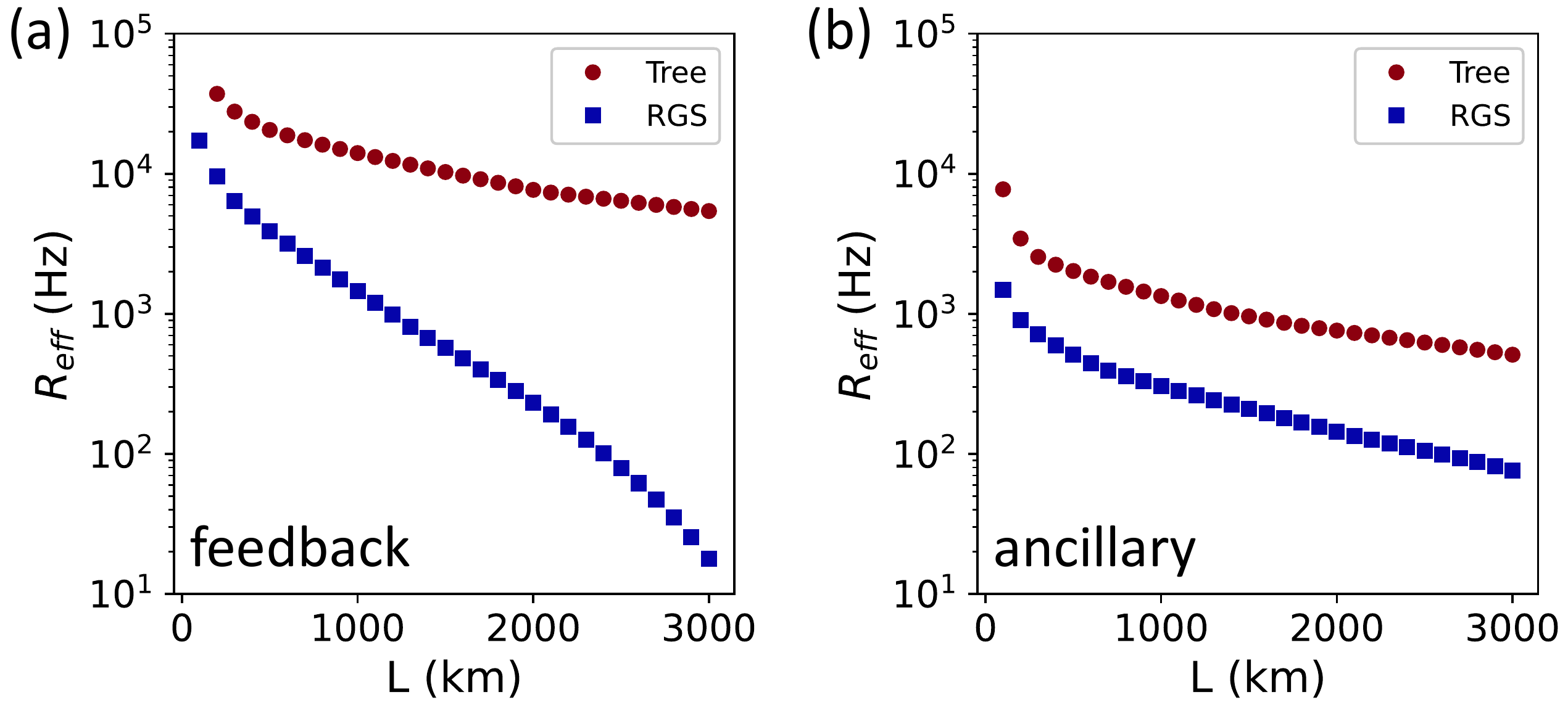}
\caption{The maximally achievable effective secret key rate $R_{\text{eff}}$ as a function of total distance between Alice and Bob, $L$, for the tree repeater (red dots) and RGS repeater (blue squares) protocols, when implemented by the feedback-assisted (a) and ancillary-qubit-assisted (b) generation schemes. Here, we assume $\gamma/2\pi=10$ GHz and $t_{\text{coh}}=1$ ms.}
\label{fig5}
\end{figure}

Finally, we compare the performance of the two repeater protocols when implemented by the same generation scheme. Figure~\ref{fig5}(a) shows the performance of the tree (red dots) and RGS (blue squares) repeater protocols as a function of distance $L$, when the protocols are implemented by the feedback-assisted scheme. In this calculation, we again fix $\gamma/2\pi=10$ GHz and $t_{\text{coh}}=1$ ms. In this case, the tree repeater protocol clearly outperforms the RGS repeater protocol and shows a slower decay with the end-to-end distance. This is because the tree repeater requires a graph state that contains an order of magnitude less photons than the RGS repeater protocol, to which the feedback-assisted scheme is sensitive. In contrast, when implemented using the ancillary-qubit-assisted scheme, as shown in Fig.~\ref{fig5}(b), the tree repeater protocol is again better than the RGS repeater protocol for all distances, but they show a similar trend as a function of distance. This is because the performance of the ancillary-qubit-assisted scheme is less sensitive to the size of the graph state.

\section{Conclusions and Discussions}
\label{conclusion}

In conclusion, we quantitatively compared the performance of two recently proposed photonic graph state generation schemes in the realization of two quantum repeater protocols based on photonic graph states. While both generation schemes rely on quantum emitters to sequentially emit photons of the graph state, the different ancillary resources required in these schemes make them advantageous in different parameter regimes. The ancillary-qubit-assisted generation scheme is better suited when the optical linewidth of the emitter is small but the spin coherence time is long. In contrast, the feedback-assisted scheme works better when the optical linewidth of the emitter is large. In addition, our results suggest that when the optical linewidth is reasonably large, the feedback-assisted scheme and tree repeater protocol generally performs better than their counterparts at all distances in the range $L<3000$ km.

Our results provide clear guidance on the selection of the numerically optimal generation scheme and protocol for graph-state-based quantum repeaters, and lay out the parameter requirements for future experimental realizations of different schemes. For example, the feedback-assisted generation scheme should be preferred for systems possessing a large oscillator strength that can exhibit a large radiative emission rate to a specific waveguide or cavity mode, such as self-assembled quantum dots~\cite{PhysRevLett.113.093603,Ota:2018ut,Najer:2019to}, silicon-vacancy (SiV) centers in diamond~\cite{Bhaskar:2020uj}, or atomic ensembles where superradiance can be employed~\cite{PhysRevLett.117.073003}. Other atomic systems, such as trapped neutral atoms~\cite{Thomas:2022tr,RevModPhys.87.1379}, rare-earth ion impurities~\cite{Kindem:2020vz,Raha:2020te}, or many quantum defect centers~\cite{Zhang:2020ua}, in general possess weaker radiative decay rates and therefore may be more suited for the ancillary-qubit-assisted scheme, especially if they can be easily coupled to nearby nuclear spins, which can serve as the ancillary matter qubits~\cite{doi:10.1126/science.1139831,PhysRevX.9.031045,Ruskuc:2022uy,doi:10.1126/science.add9771}.

Our analysis can be easily extended to study other generation schemes and other types of all-photonic quantum repeater protocols~\cite{PhysRevLett.117.210501,PhysRevA.100.052303,PhysRevA.104.052623,Zhang:22}, such as the recently proposed protocol based on logical Bell state measurements using tree graph states, which is much more resistant to logical errors besides the photon loss correction~\cite{PhysRevA.104.052623}. It can also be used to study how some of the other realistic imperfections impact the overall performance of a graph-state-based quantum repeater, such as the imperfect fidelity of each gate operation, which was assumed to be perfect in our analysis. It is interesting to note that non-ideal gate operations will lead to single-photon logical errors. It remains an open question how these logical errors will propagate under different generation schemes, and whether one scheme is superior to the other in correcting such errors~\cite{PRXQuantum.2.040345}. Overall, our results represent an important step forward towards the development of graph-state-based quantum repeaters and the quantum internet.

\section*{Acknowledgements}
We acknowledge funding from National Science Foundation (Grant No. 1734006, 2016244, and 2137953). This work utilized resources from the University of Colorado Boulder Research Computing Group, which is supported by the National Science Foundation (awards ACI-1532235 and ACI-1532236), the University of Colorado Boulder, and Colorado State University. Y.Z. acknowledges support from NSF QISE-NET award (funded by NSF Grant No. 1747426). S.E.E. acknowledges support from the Army Research Office (MURI Grant No. W911NF2120214).

\bibliographystyle{elsarticle-num}
\bibliography{main_quantum}

\onecolumn
\newpage
\appendix

\section{Feedback-assisted generation scheme: a modified version from Ref.~\cite{PhysRevLett.125.223601}}
\label{appendixA}

The feedback-assisted generation scheme discussed in the main text is based on a modified version of Ref.~\cite{PhysRevLett.125.223601}. The modification is made in order to improve the generation speed and the error robustness. For the generation of tree graph states, in the original scheme shown in Ref.~\cite{PhysRevLett.125.223601}, all the photons will interact with the emitter through a scattering process upon being routed back from the feedback line. We can simplify the generation process by generating the bottom layer photons using the $P$ gate. In this way, the bottom layer photons will be entangled with the spin qubit of the quantum emitter upon emission, and therefore do not need to go through the feedback line. This modification allows us to shrink the length of the required feedback line, thereby reducing internal photon losses in the generation process.

For the generation of repeater graph states with tree encoding, the original proposal starts with the generation of a complex tree state, followed by single-photon $\sigma_X$ measurements on certain photons in the tree state. However, this generation process is only successful upon the survival of these photons, and the fidelity of the generated state using this method is extremely sensitive to errors in each single-photon $\sigma_X$ measurement. Therefore, we have adapted a new generation scheme for repeater graph states as shown in Fig.~\ref{fig2}(d) of the main text. The new generation scheme does not require any single-photon measurements, but requires all the level-1 photons in the encoding tree graph of each core node to scatter twice from the emitter. 

\begin{figure}[!t]
\centering
\includegraphics[width=0.7\columnwidth]{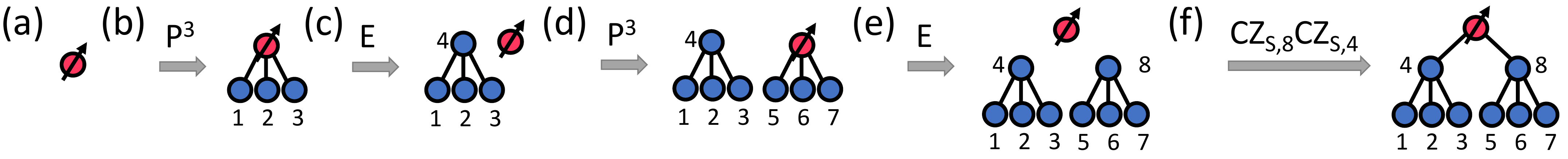}
\caption{Graph representation of the procedure for generating an example tree graph state with branching parameters $\{b_0=2,b_1=3\}$. (a) The spin of the quantum emitter is initially prepared in state $\ket{+}_s=\frac{1}{\sqrt{2}}\left(\ket{0}_s+\ket{1}_s\right)$. (b) Sequentially applying the $P$ gate for 3 times generates photons 1-3 in the bottom layer, which are entangled with the spin of the quantum emitter. (c) Applying an $E$ gate generates photon 4, which inherits the entanglement of the emitter spin and detaches the emitter spin from the graph. (d, e) Repeating the same procedure shown in (b) and (c) generates another small tree which consists of photons 5-8. (f) Applying two $CZ$ gates entangles the emitter spin with photons 4 and 8, which completes the whole tree with the root node being the emitter spin. The indices of photons indicate the generation order.}
\label{afig1}
\bigskip
% \end{figure}

% \begin{figure}[t]
% \centering
\includegraphics[width=0.75\columnwidth]{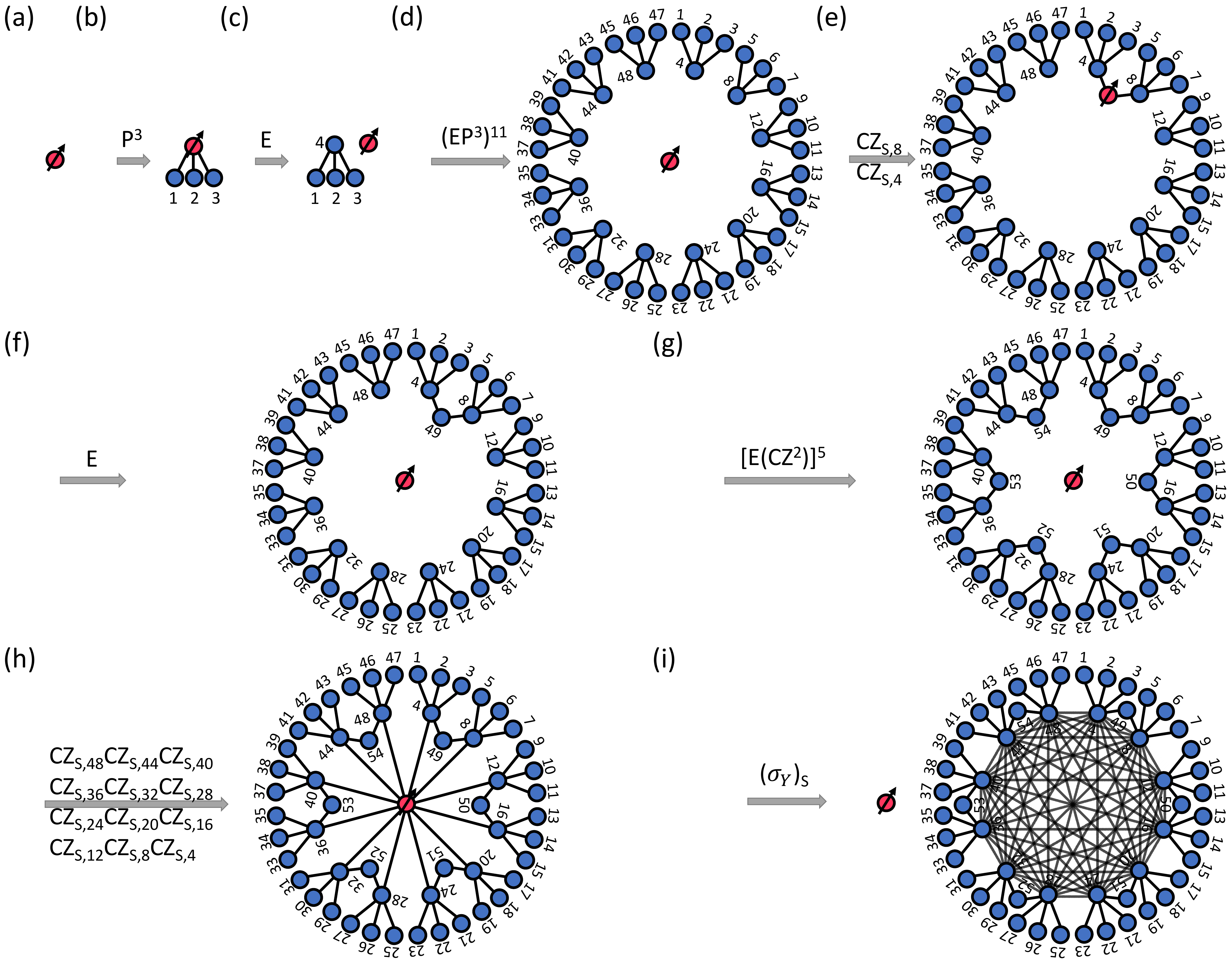}
\caption{Graph representation of the procedure for generating an example repeater graph state with 6 core nodes and 6 arm nodes. Each core node is encoded by a tree graph with branching parameters $\{b_0=2,b_1=3\}$. (a) The spin of the quantum emitter is initially prepared in state $\ket{+}_s=\frac{1}{\sqrt{2}}\left(\ket{0}_s+\ket{1}_s\right)$. (b) Sequentially applying the $P$ gate for 3 times generates photons 1-3, which are entangled with the spin of the quantum emitter. (c) Applying an $E$ gate generates photon 4, which inherits the entanglement of the emitter spin and detaches the emitter spin from the graph. (d) Repeating the same procedure shown in (b) and (c) for 11 times generates 11 more small trees. (e) Applying two $CZ$ gates entangles the emitter spin with photons 4 and 8. (f) Applying an $E$ gate generates photon 49 which inherits the entanglement of the emitter spin with photons 4 and 8, and detaches the emitter spin from the graph. (g) Repeating the same procedure shown in (e) and (f) for 5 more times generates 5 more larger trees. (h) Applying 6 $CZ$ gates between the emitter spin and all level-1 photons of the larger trees again entangles the emitter spin with them. (i) Applying a single-qubit measurement in the $\sigma_Y$ basis on the emitter spin performs a local complementation operation which fully connects all level-1 photons and completes the generation process. The indices of photons indicate the generation order.}
\label{afig2}
\end{figure}

Figures~\ref{afig1} and~\ref{afig2} illustrate the processes of generating the tree and repeater graph states, respectively, based on the modified feedback-assisted generation scheme.

\section{Calculation of the quantum state fidelity}
\label{appendixB}

The quantum state fidelity of interest, denoted as $F$, has different definitions in different repeater protocols. In the tree repeater, $F$ is defined as the quantum state fidelity of the transmitted logical qubit from Alice to Bob. In the RGS repeater, $F$ is defined as the quantum state fidelity of the shared Bell state between Alice and Bob. To explicitly derive $F$ for both repeater protocols, we consider a similar error model as in Ref.~\cite{Azuma:2015vq,PhysRevX.10.021071,Hilaire2021resource}, and assume an independent depolarization channel for each photon parameterized by $\varepsilon$,
\begin{equation}
\begin{aligned}
\Delta_\varepsilon(\rho)=(1-\varepsilon)\rho+\frac{\varepsilon}{3}\left(\sigma_X\rho\sigma_X+\sigma_Y\rho\sigma_Y+\sigma_Z\rho\sigma_Z\right),
\label{aeq1}
\end{aligned}
\end{equation}
where $\Delta_\varepsilon$ is the depolarization operator, $\rho$ is the density matrix of a single photon, and $\sigma_{X,Y,Z}$ are Pauli matrices. Therefore, the probability that we get an incorrect single-photon measurement outcome under any measurement basis is given by $\varepsilon_\text{sp}=\frac{2}{3}\varepsilon$. A detailed analysis of the dependence of $F$ on $\varepsilon_\text{sp}$ for the RGS repeater has been performed in the Supplementary Information of Ref.~\cite{Azuma:2015vq}, but the analytical derivation of $F$ for the tree repeater is missing. Here, for the first time, we derive the relationship between $F$ and $\varepsilon_\text{sp}$ for the tree repeater.

In the tree repeater protocol, $F$ can be expressed as
\begin{equation}
\begin{aligned}
F=(1-\bar{e}_\text{decoding})^{m+1},
\label{aeq2}
\end{aligned}
\end{equation}
where $m$ is the number of repeater nodes between Alice and Bob, and $\bar{e}_\text{decoding}$ is the conditional error probability of the decoded logical qubit at each repeater node given that a result is successfully achieved, which is given by
\begin{equation}
\begin{aligned}
\bar{e}_\text{decoding}=\frac{\bar{e}_\text{incorrect}}{P_\text{succ}^\text{tree}},
\label{aeq3}
\end{aligned}
\end{equation}
where $P_\text{succ}^\text{tree}$ is given by Eq.~\ref{eq3} in the main text, and $\bar{e}_\text{incorrect}$ is the average probability of decoding an incorrect result at each repeater node. In order to successfully decode a result (either correct or incorrect) at each repeater node, we need to measure all the surviving photons in the tree graph state in either the $\sigma_X$ or $\sigma_Z$ basis. Specifically, one photon in the first level of the tree, which is entangled with the matter qubit, needs to be measured under the $\sigma_X$ basis, whereas all its leaf photons in the second level, together with all other photons in the first level, need to be measured under the $\sigma_Z$ basis, either directly or indirectly. A correct decoded result requires the following three conditions to be satisfied simultaneously: (i) the first-level photon measured under the $\sigma_X$ basis needs to be errorless; (ii) the parity of all the $\sigma_Z$ measurement outcomes from all other first-level photons, either direct or indirect, needs to be errorless; (iii) the parity of all the $\sigma_Z$ measurement outcomes from all the second-level photons that are entangled with the $\sigma_X$-measured first-level photon, either direct or indirect, needs to be errorless. Therefore, $\bar{e}_\text{incorrect}$ can be estimated as
\begin{equation}
\begin{aligned}
\bar{e}_\text{incorrect}=\sum_{l=0}^{b_0-1}\sum_{n=0}^{b_0-l}\sum_{m=0}^{b_1}&\left(
        \begin{array}{c}
        b_0 \\
        l
        \end{array}
\right)\left(
        \begin{array}{c}
        b_0-l \\
        n
        \end{array}
\right)\left(
        \begin{array}{c}
        b_1 \\
        m
        \end{array}
\right) \\
&\times(\mu R_1)^l[(1-\mu)(1-R_1)]^n [(1-\mu)R_1]^{b_0-l-n}[(1-\mu)(1-R_2)]^m R_2^{b_1-m}\bar{e}_{n,m},
\label{aeq4}
\end{aligned}
\end{equation}
where $\mu$ is the single-photon loss probability between neighboring repeater nodes, $R_i$ is the probability of obtaining an indirect $\sigma_Z$ measurement outcome on any given photon found in the $i$-th level of the tree graph state. In Eq.~\ref{aeq4}, $l$ is the number of photons situated in the first level of the tree that are physically lost but the indirect $\sigma_Z$ measurements on them can succeed, $n$ is the number of photons situated in the first level of the tree that survive but the indirect $\sigma_Z$ measurements on them will fail, and $(b_0-l-n)$ is the number of photons situated in the first level of the tree that survive and the indirect $\sigma_Z$ measurements on them can succeed. Note that we adopt a strategy that if both direct and indirect $\sigma_Z$ measurements can be performed on a photon, we choose the indirect result for decoding since the error correction capability of the tree encoding should give a smaller error probability. Therefore, one photon out of the $(b_0-l)$ surviving first-level photons will be used for entangling with the matter qubit and the $\sigma_X$ measurement for decoding, $n$ first-level photons will provide the direct $\sigma_Z$ measurement outcomes, and $(b_0-n-1)$ first-level photons will provide the indirect $\sigma_Z$ measurement outcomes. In Eq.~\ref{aeq4}, similarly, for the $b_1$ second-level photons that are entangled with the $\sigma_X$-measured first-level photon, $m$ is the number of photons that survive but the indirect $\sigma_Z$ measurements on them fail, and $(b_1-m)$ is the number of photons on which the indirect $\sigma_Z$ measurements succeed (they can be physically lost or not lost). In Eq.~\ref{aeq4}, $\bar{e}_{n,m}$ is the average error probability of the decoded logical qubit in the case in which $n$ measurement outcomes under the $\sigma_Z$ basis from the first level and $m$ measurement outcomes under the $\sigma_Z$ basis in the second level of the tree are from the direct measurements, while all other $\sigma_Z$ measurement outcomes are obtained from the indirect measurements. $\bar{e}_{n,m}$ can be estimated as
\begin{equation}
\begin{aligned}
\bar{e}_{n,m}=1-(1-\varepsilon_\text{sp})&\left\{1-\sum_{i=0}^n\left[\left(\begin{array}{c}n \\ i\end{array}\right)\varepsilon_\text{sp}^i(1-\varepsilon_\text{sp})^{n-i}\sum_{\substack{j=0,\\ i+j=1[2]}}^{b_0-1-n}\left(\begin{array}{c}b_0-1-n \\ j\end{array}\right)\bar{e}_{I_1}^j(1-\bar{e}_{I_1})^{b_0-1-n-j}\right]\right\} \\
&\times\left\{1-\sum_{i=0}^m\left[\left(\begin{array}{c}m \\ i\end{array}\right)\varepsilon_\text{sp}^i(1-\varepsilon_\text{sp})^{m-i}\sum_{\substack{j=0,\\ i+j=1[2]}}^{b_1-m}\left(\begin{array}{c}b_1-m \\ j\end{array}\right)\bar{e}_{I_2}^j(1-\bar{e}_{I_2})^{b_1-m-j}\right]\right\},
\label{aeq5}
\end{aligned}
\end{equation}
where $\bar{e}_{I_k}$ is the average error probability of guessing the indirect $\sigma_Z$ measurement outcome on a level-$k$ photon from measuring one of its branches, which can be estimated using a majority vote strategy, as explicitly calculated in the Supplementary Information in Ref.~\cite{Azuma:2015vq}. We refer the readers to Ref.~\cite{Azuma:2015vq} for its detailed derivation.

We now relate $\varepsilon_\text{sp}$ to concrete experimental parameters we consider in the main text: the spin coherence time of the quantum emitter $t_\text{coh}$ and the depolarization probability of each photon in the communication channel $\varepsilon_\text{depol}$. We model the spin decoherence process as a depolarization channel parameterized by $\varepsilon_\text{decoh}$ on each photon of the tree graph state, which can be described by
\begin{equation}
\begin{aligned}
\Delta_{\varepsilon_\text{decoh}}(\rho)=(1-\varepsilon_\text{decoh})\rho+\frac{\varepsilon_\text{decoh}}{3}\left(\sigma_X\rho\sigma_X+\sigma_Y\rho\sigma_Y+\sigma_Z\rho\sigma_Z\right),
\label{aeq6}
\end{aligned}
\end{equation}
where $\varepsilon_\text{decoh}$ can be expressed as
\begin{equation}
\begin{aligned}
\varepsilon_\text{decoh}=\frac{3}{4}\left[1-\exp{\left(\frac{T_\text{graph}}{t_\text{coh}N_\text{ph}}\right)}\right],
\end{aligned}
\label{aeq7}
\end{equation}
where $T_\text{graph}$ is the time it takes for each repeater node to generate the required photonic graph state, $t_\text{coh}$ is the spin coherence time of the quantum emitter, and $N_\text{ph}$ is the total number of photons in the graph state.

We model the photon depolarization process in the communication channel as
\begin{equation}
\begin{aligned}
\Delta_{\varepsilon_\text{depol}}(\rho)=(1-\varepsilon_\text{depol})\rho+\frac{\varepsilon_\text{depol}}{3}\left(\sigma_X\rho\sigma_X+\sigma_Y\rho\sigma_Y+\sigma_Z\rho\sigma_Z\right).
\label{aeq8}
\end{aligned}
\end{equation}

These two error processes can be combined into one effective depolarization channel described by Eq.~\ref{aeq1}, therefore mapping $\varepsilon_\text{decoh}$ and $\varepsilon_\text{depol}$ into $\varepsilon_\text{sp}$ as
\begin{equation}
\begin{aligned}
\varepsilon_\text{sp}=\frac{2}{3}\left(\varepsilon_\text{decoh}+\varepsilon_\text{depol}-\frac{4}{3}\varepsilon_\text{decoh}\varepsilon_\text{depol}\right).
\label{aeq9}
\end{aligned}
\end{equation}

\section{Calculation of the length of the feedback line in the feedback-assisted generation scheme}
\label{appendixC}

To generate a tree graph state with branching parameters $\{b_i\}$ and depth $d$, the minimum required length of the feedback line in the feedback-assisted generation scheme is given by
\begin{equation}
\begin{aligned}
L_{\text{feedback}}^{\text{tree}}=[(n_{d-1}+n_{d-2}-1)t_E^f+b_{d-1}(n_{d-1}-n_{d-2})t_P^f]v_{\text{feedback}},
\label{aeq10}
\end{aligned}
\end{equation}
where $n_l=\sum_{i=0}^{l-1}b_i$ is the total number of nodes in the $l$-th level of the tree graph, $t_E^f$ and $t_P^f$ are the operation times of an $E$ gate and a $P$ gate in the feedback-assisted scheme, respectively, and $v_{\text{feedback}}$ is the velocity of light in the feedback line.

To generate a repeater graph state with $N$ core nodes and $N$ arm nodes, where each core node is encoded by a tree graph with branching parameters $\{b_i\}$ and depth $d$, the minimum required feedback line length is given by
\begin{equation}
\begin{aligned}
L_{\text{feedback}}^{\text{RGS}}=N[(n_{d-1}+n_{d-2}-1)t_E^f+b_{d-1}(n_{d-1}-n_{d-2})t_P^f]v_{\text{feedback}},
\label{aeq11}
\end{aligned}
\end{equation}
where $n_l$ is the total number of photons in the $l$-th level of the encoding tree graph state.

\section{Calculation of the length of the delay line to correct photon arrival order}
\label{appendixD}

\begin{figure}[t]
\centering
\includegraphics[width=0.55\columnwidth]{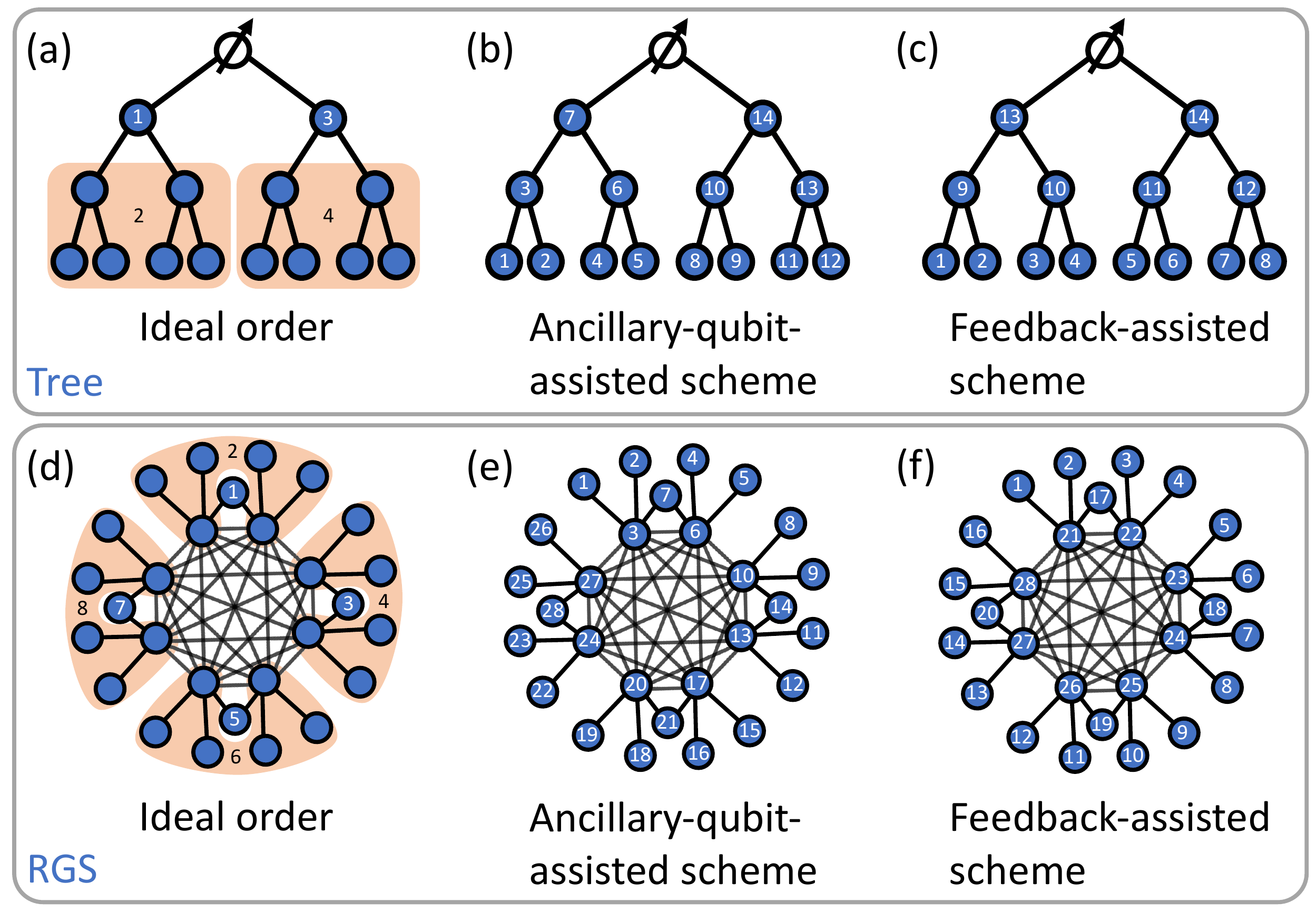}
\caption{(a)-(c) Ideal order for single-photon measurements at each repeater node (a), the output photon order in the ancillary-qubit-assisted scheme (b), and the output photon order in the feedback-assisted scheme (c) in the tree repeater protocol, presented with a tree graph state with branching parameters $\{b_0=b_1=b_2=2\}$. (d)-(f) Ideal order for single-photon measurements at each measurement node (d), the output photon order in the ancillary-qubit-assisted scheme (e), and the output photon order in the feedback-assisted scheme (f) in the RGS repeater protocol, presented with a repeater graph state with $N=4$ branches where each core qubit is encoded by a tree graph state $\{b_0=b_1=2\}$. The indices indicate the corresponding orders. The orange shadows in (a) and (d) mean that the measurement order of photons inside the same shadow does not matter. Note that the photon output order in the feedback-assisted scheme [(c) and (f)] is not the same as their emission order presented in Fig.~\ref{afig1} and Fig.~\ref{afig2}. This is because different photons in the graph state generated by the feedback-assisted scheme may need to travel through different numbers of rounds trips in the feedback line before they are output to the communication channel.}
\label{afig3}
\end{figure}

The photons in either a tree or a repeater graph state generated by either the ancillary-qubit- or feedback-assisted scheme are not in the desired order for measurements. Therefore a delay line is required before they are measured, whose length depends on both the repeater protocol and the generation scheme. In the tree repeater, each first level photon should be measured before all other photons that are situated in the same branch~\cite{PhysRevX.10.021071}, as shown in Fig.~\ref{afig3}(a). However, in the realization with the ancillary-qubit-assisted scheme, photons are sent out to the communication channel from bottom to top in each branch of the tree [Fig.~\ref{afig3}(b)], whereas in the realization with the feedback-assisted scheme, photons are sent out to the communication channel from bottom to top in the whole tree [Fig.~\ref{afig3}(c)]. We thus estimate the delay line lengths for the two generations schemes in the tree repeater to be
\begin{equation}
\begin{aligned}
L_{\text{delay}}^{\text{tree,ancilla}}&\approx \left[\prod_{i=1}^{d-1}b_it_P^a+\left(\beta+\sum_{l=1}^{d-2}\prod_{i=1}^{l}b_i\right)t_E^a+\sum_{l=1}^{d-2}\prod_{i=0}^{l}b_it_{\text{CZ}}^a\right]v_{\text{delay}}, \\
L_{\text{delay}}^{\text{tree,feedback}}&\approx \left(d-1+\frac{1}{b_0}\right)L_{\text{feedback}}^{\text{tree}},
\label{aeq12}
\end{aligned}
\end{equation}
where $b_i$ and $d$ are the branching parameters and the depth of the tree graph state, $t_{P,E,CZ}^a$ and $\beta$ have the same definitions as in the main text, $v_{\text{delay}}$ is the velocity of light in the delay line, and $L_{\text{feedback}}^{\text{tree}}$ is the length of the required feedback line during the generation process, which is given by Eq.~\ref{aeq10} in Appendix~\ref{appendixC}.

In the RGS repeater, each arm photon should be measured before all other photons that are situated in the same branch~\cite{Azuma:2015vq}, as shown in Fig.~\ref{afig3}(d). However, in the realization with the ancillary-qubit-assisted scheme, photons are sent out to the communication channel from bottom to top in each branch of the RGS [Fig.~\ref{afig3}(e)]. In the realization with the feedback-assisted scheme, photons in levels larger than 1 in the encoding tree graph state are sent to the communication channel first, then the arm photons, and last the first level photons [Fig.~\ref{afig3}(f)]. We thus estimate the delay line lengths for the two generations schemes in the RGS repeater to be
\begin{equation}
\begin{aligned}
L_{\text{delay}}^{\text{RGS,ancilla}}&\approx \left[\left(1+\prod_{i=0}^{d-1}b_i\right)t_P^a+\sum_{l=0}^{d-2}\prod_{i=0}^{l}b_i\left(t_E^a+t_{CZ}^a\right)\right]v_{\text{delay}}, \\
L_{\text{delay}}^{\text{RGS,feedback}}&\approx \left(d-1+\frac{1}{N}\right)L_{\text{feedback}}^{\text{RGS}},
\label{aeq13}
\end{aligned}
\end{equation}
where $b_i$ and $d$ are the branching parameters and the depth of the encoding tree graph state, and $L_{\text{feedback}}^{\text{RGS}}$ is the length of the required feedback line during the generation process, which is given by Eq.~\ref{aeq11} in Appendix~\ref{appendixC}.

\section{Ratio of wavepacket length between long photons and the rest}
\label{appendixE}

In the application of the ancillary-qubit-assisted scheme in the tree repeater protocol, along with the application of the feedback-assisted scheme in both repeater protocols, some photons in the graph states are required to have longer wavepackets than other photons in order to boost the fidelity of the spin-photon $CZ$ gate. The ratio of the wavepacket length between these longer photons and the rest of the photons is defined as $\beta$. We choose $\beta=500$ in the main text so that the infidelity of the spin-photon $CZ$ gate, denoted as $\varepsilon_\text{CZ}$, is negligible compared with other error sources. Specifically, as calculated in the Supplementary Material of Ref.~\cite{PhysRevLett.125.223601}, under the assumption of a Gaussian wavepacket for the longer photons, $\varepsilon_\text{CZ}$ can be given by
\begin{equation}
\begin{aligned}
\varepsilon_\text{CZ}=\frac{2}{\beta^2}+\mathcal{O}\left(\frac{1}{\beta^4}\right).
\label{aeq14}
\end{aligned}
\end{equation}
For $\beta=500$, this gives $\varepsilon_\text{CZ}=8\times 10^{-6}$. This is one order of magnitude smaller than the depolarization probability of each photon in the communication channel, $\varepsilon_\text{depol}$, which we assume to be $5\times 10^{-5}$.

\section{Numerically optimized quantum repeater setup for specific quantum emitter parameters}
\label{appendixF}

\begin{table}[t]
\caption{Numerically optimized quantum repeater setup for specific $\gamma$ and $t_\text{coh}$ for both repeater protocols and generation schemes at $L=1000$ km.}
\begin{threeparttable}
%\begin{ruledtabular}
\begin{tabularx}{1.0\columnwidth}{p{0.12\columnwidth}p{0.06\columnwidth}p{0.17\columnwidth}p{0.17\columnwidth}p{0.17\columnwidth}p{0.17\columnwidth}}
\hline\hline
$\gamma$ &$t_{\text{coh}}$ &Tree repeater with ancillary-qubit-assisted scheme &Tree repeater with feedback-assisted scheme &RGS repeater with ancillary-qubit-assisted scheme &RGS repeater with feedback-assisted scheme \\
\hline
$2\pi\times 2$ GHz &13 ms &\begin{tabular}{l}$R_\text{eff}$: 1.4 kHz \\ $\frac{L}{m+1}$: 1.7 km \\ $\{b_i\}$: $\{4, 16, 5\}$ \\ $L_\text{delay}$: 398 m  \end{tabular} &\begin{tabular}{l}$R_\text{eff}$: 817.4 Hz \\ $\frac{L}{m+1}$: 1.1 km \\ $\{b_i\}$: $\{4, 16, 5\}$ \\ $L_\text{feedback}$: 540.3 m\\ $L_\text{delay}$: 1.2 km  \end{tabular} &\begin{tabular}{l}$R_\text{eff}$: 321.5 Hz \\ $\frac{L}{m+1}$: 3.2 km \\ $N$: 32 \\ $\{b_i\}$: $\{24,7\}$ \\ $L_\text{delay}$: 483.5 m \end{tabular} &\begin{tabular}{l}$R_\text{eff}$: $2.1\times 10^{-7}$ Hz \\ $\frac{L}{m+1}$: 2.2 km \\ $N$: 14 \\ $\{b_i\}$: $\{13,5\}$ \\ $L_\text{feedback}$: 1.5 km \\ $L_\text{delay}$: 1.6 km  \end{tabular} \\ \hline
$2\pi\times 100$ GHz &4 $\mu$s &\begin{tabular}{l}$R_\text{eff}$: $2.5\times 10^{-17}$ Hz  \\ $\frac{L}{m+1}$: 30.3 km \\ $\{b_i\}$: $\{1,1,19\}$ \\ $L_\text{delay}$: 80.4 m \end{tabular} &\begin{tabular}{l}$R_\text{eff}$: 104.8 kHz \\ $\frac{L}{m+1}$: 2.2 km \\ $\{b_i\}$: $\{4,22,6\}$ \\ $L_\text{feedback}$: 16.5 m \\ $L_\text{delay}$: 37.1 m \end{tabular} &\begin{tabular}{l}$R_\text{eff}$: $7.3\times 10^{-80}$ Hz \\ $\frac{L}{m+1}$: 500 km \\ $N$: 4 \\ $\{b_i\}$: $\{1,1\}$ \\ $L_\text{delay}$: 20.0 m \end{tabular} &\begin{tabular}{l}$R_\text{eff}$: 33.3 kHz \\ $\frac{L}{m+1}$: 3.9 km \\ $N$: 32 \\ $\{b_i\}$: $\{25,7\}$ \\ $L_\text{feedback}$: 145.1 m \\ $L_\text{delay}$: 149.6 m  \end{tabular} \\ \hline
$2\pi\times 170$ MHz &1 s &\begin{tabular}{l}$R_\text{eff}$: 988.0 Hz \\ $\frac{L}{m+1}$: 1.8 km \\ $\{b_i\}$: $\{4,18,5\}$ \\ $L_\text{delay}$: 627.9 m \end{tabular} &\begin{tabular}{l}$R_\text{eff}$: $7.4\times 10^{-18}$ Hz \\ $\frac{L}{m+1}$: 27.0 km \\ $\{b_i\}$: $\{1,1,18\}$ \\ $L_\text{feedback}$: 93.6 m \\ $L_\text{delay}$: 378.0 m \end{tabular} &\begin{tabular}{l}$R_\text{eff}$: 293.9 Hz \\ $\frac{L}{m+1}$: 3.1 km \\ $N$: 32 \\ $\{b_i\}$: $\{24,7\}$ \\ $L_\text{delay}$: 516.6 m \end{tabular} &\begin{tabular}{l}$R_\text{eff}$: $1.8\times 10^{-68}$ Hz \\ $\frac{L}{m+1}$: 9.3 km \\ $N$: 4 \\ $\{b_i\}$: $\{4,2\}$ \\ $L_\text{feedback}$: 1.5 km \\ $L_\text{delay}$: 1.9 km \end{tabular} \\ \hline
$2\pi\times 100$ GHz &1 s &\begin{tabular}{l}$R_\text{eff}$: 1.5 kHz \\ $\frac{L}{m+1}$: 1.7 km \\ $\{b_i\}$: $\{4,16,5\}$ \\ $L_\text{delay}$: 380.7 m \end{tabular} &\begin{tabular}{l}$R_\text{eff}$: 151.1 kHz \\ $\frac{L}{m+1}$: 1.9 km \\ $\{b_i\}$: $\{4,15,5\}$ \\ $L_\text{feedback}$: 11.3 m \\ $L_\text{delay}$: 25.6 m \end{tabular} &\begin{tabular}{l}$R_\text{eff}$: 326.1 Hz \\ $\frac{L}{m+1}$: 3.2 km \\ $N$: 32 \\ $\{b_i\}$: $\{24,7\}$ \\ $L_\text{delay}$: 480.5 m \end{tabular} &\begin{tabular}{l}$R_\text{eff}$: 48.7 kHz \\ $\frac{L}{m+1}$: 3.8 km \\ $N$: 32 \\ $\{b_i\}$: $\{24,7\}$ \\ $L_\text{feedback}$: 139.3 m \\ $L_\text{delay}$: 143.7 m \end{tabular} \\
\hline\hline
\end{tabularx}
%\end{ruledtabular}
\begin{tablenotes}
\footnotesize
\item $R_\text{eff}$: The maximally achievable effective secret key rate;
\item $\frac{L}{m+1}$: The distance between neighboring repeater nodes in the tree repeater, or the distance between neighboring source nodes in the RGS repeater, where $L=1000$ km;
\item $N$: The number of branches in the numerically optimized repeater graph state in the RGS repeater;
\item $b_i$: The branching parameters of the numerically optimized tree graph state in the tree repeater, or the numerically optimized encoding tree graph state in the RGS repeater;
\item $L_\text{feedback}$: The length of the feedback line in the feedback-assisted generation scheme, calculated in Appendix~\ref{appendixC};
\item $L_\text{delay}$: The length of the delay line for photon arrival order correction, calculated in Appendix~\ref{appendixD}.
\end{tablenotes}
\end{threeparttable}
\label{atab1}
\end{table}

Figure~\ref{fig3} in the main text shows the maximally achievable effective key rate $R_{\text{eff}}$ for both repeater protocols realized by both generation schemes as a function of the optical linewidth $\gamma$ and the spin coherence time $t_{\text{coh}}$ of the quantum emitter. Here we present the numerically optimized quantum repeater setup (including the numerically optimized photonic graph state shape, the distance between neighboring repeater nodes, and the length of the feedback or delay line) for several specific parameters $\gamma$ and $t_{\text{coh}}$, as shown in Table~\ref{atab1}. These chosen parameters correspond to typical physical systems including a single silicon-vacancy (SiV) color center coupled with a photonic crystal cavity ($\gamma\sim 2\pi\times 2$ GHz~\cite{Bhaskar:2020uj} and $t_\text{coh}\sim 13$ ms~\cite{PhysRevLett.119.223602}), a single semiconductor quantum dot coupled with a nano-cavity ($\gamma\sim 2\pi\times 100$ GHz~\cite{Ota:2018ut} and $t_\text{coh}\sim 4$ $\mu$s~\cite{PhysRevB.97.241413}), and a single neutral atom coupled with a fiber Fabry-Perot cavity ($\gamma\sim 2\pi\times 170$ MHz~\cite{Brekenfeld:2020um} and $t_\text{coh}\sim 1$ s~\cite{RevModPhys.87.1379}). Nitrogen-vacancy (NV) centers in diamond, which have been demonstrated to have a long electron spin coherence time of $\sim 1$ s~\cite{Abobeih:2018td}, are also possible platforms if significant enhancement of both the zero phonon line fraction and the emission rate can be achieved when coupled to a micro-~\cite{PhysRevX.7.031040} or nano-cavity~\cite{Jung:2019tu}. These systems are all promising candidates for the experimental realizations of the ultrafast quantum repeaters based on photonic graph states upon reasonable improvements. We also present the numerically optimized repeater setup for an ideal quantum emitter with parameters $\gamma=2\pi\times 100$ GHz and $t_\text{coh}=1$ s, which combines the best values from the physical systems mentioned above.

\end{document}